\documentclass[a4paper,12pt]{article}
\usepackage[utf8x]{inputenc}
\usepackage{amsmath,amssymb,graphicx,a4wide}
\newcommand{\abs}[1]{\lvert #1\rvert}
\newcommand{\avg}[1]{\langle #1\rangle}
\newcommand{\bavg}[1]{\big\langle #1\big\rangle}
\newcommand{\onepic}[1]{\includegraphics[scale=0.25]{#1}}
\newcommand{\mymod}{\%}
\newcommand{\dd}{\mathrm{d}}

\newcommand{\ee}{\mathrm{e}}
\begin{document}
\title{Transaction fees and optimal rebalancing\\
in the growth-optimal portfolio}
\author{Yu Feng$^{1,2}$, Matúš Medo$^3$, Liang Zhang$^4$,
Yi-Cheng Zhang$^{1,2,3,4}$}
\date{}
\maketitle

{\small\noindent
$^1$ Physics Department, Renmin University, 100872
Beijing, PR China\\
$^2$ Complexity Research Center, USST, 200093 Shanghai,
PR China\\
$^3$ Physics Department, University of Fribourg,
1700~Fribourg, Switzerland\\
$^4$ Lab of Information Economy and Internet Research,
University of Electronic Science\\
\phantom{$^4$ }and Technology, 610054 Chengdu, PR China}

\begin{abstract}
The growth-optimal portfolio optimization strategy pioneered by
Kelly is based on constant portfolio rebalancing which makes it
sensitive to transaction fees. We examine the effect of fees on
an example of a risky asset with a binary return distribution
and show that the fees may give rise to an optimal period of
portfolio rebalancing. The optimal period is found analytically
in the case of lognormal returns. This result is consequently
generalized and numerically verified for broad return
distributions and returns generated by a GARCH process. Finally
we study the case when investment is rebalanced only partially
and show that this strategy can improve the investment long-term
growth rate more than optimization of the rebalancing
period.\\[12pt]
Keywords: growth-optimal portfolio, Kelly game, transaction fees,
lognormal distribution.
\end{abstract}

\section{Introduction}
Portfolio optimization is one of the main topics in quantitative
finance. The aim is to maximize investment return while
simultaneously minimizing its risk (see \cite{EGBG06,ZZ06} for a
review of the modern portfolio theory). Pioneering works on this
problem were mainly focused on the Mean-Variance
approach~\cite{Mark52} where the portfolio variance is minimized
under the constraint of a fixed expected return value. A
different approach has been put forward by Kelly~\cite{Kelly56}
who focused on repeated investments and proposed to maximize the
long-term growth rate of the investor's capital. This so-called
growth-optimal or Kelly portfolio has been shown to be optimal
according to various criteria~\cite{Mark76} and generalized in
different ways.  For example, the question of diversification
and constant rebalancing among a certain number of uncorrelated
stocks was investigated in~\cite{MaMaZh98}. In~\cite{LaMeZh10},
the authors showed that there is a close connection between the
Mean-Variance approach and the Kelly portfolio and that in many
cases, the Kelly-optimal portfolio includes only a small
fraction of the available profitable assets. When investing in
games without specified levels of risk and reward, the Kelly
criterion can be merged with Bayesian statistical learning as
in, for example, \cite{BrWh96,MePiZh08}, yielding generalized
results for the optimal investment fractions. Stochastic
portfolio theory~\cite{FerKar09} is also a descendant of Kelly's
approach by utilizing on a logarithmic representation of price
processes.

Application of Kelly's optimization process to real stock prices
was studied in~\cite{Slanina99} with the conclusion that
non-trivial investment (\emph{i.e.}, investing only a part of
one's wealth) occurs rarely. This is related to the general
notion that Kelly's portfolio is very aggressive and investment
outcomes are sensitive to errors in estimates of assets'
properties. Modifications such as fractional Kelly
strategies~\cite{MLZiBl92} and controlled
downturns~\cite{GroZho93} have been consequently proposed to
make the resulting portfolios more secure (these modifications
can be of particular importance for risky
assets~\cite{Thorp06}). Optimization in the long-term can even
explain the emergence of cooperation in environments where
outcomes of the participants are of multiplicative
nature~\cite{YaSo09}. An interested reader is referred to
\cite{Thorp06,MLeZiem06} for a comprehensive introduction to the
Kelly portfolio.

Kelly's optimization scheme is based on the long-term prospects
of the investor and requires continual rebalancing of the
portfolio which ensures that the investment fraction is kept
constant. This rebalancing represents the key advantage of the
Kelly portfolio over the simple buy-and-hold strategy. On the
other hand, when non-zero transaction costs are imposed,
resulting investment performance may deteriorate considerably
(for an example of how the transaction costs influence real
traders and their decisions see~\cite{MoCha10}). In this paper
we intend to study the effect of non-zero transaction costs on
the Kelly portfolio. We study the situation where portfolio is
rebalanced less often (intermittent rebalancing). Our key
quantity of interest is the optimal rebalancing period which
minimizes the negative effects of transaction fees while
maintaining the positive effects of frequent rebalancing.

Another reason for intermittent rebalancing is that the
distribution of returns may differ from one turn to another. We
approach this problem by postulating a risky asset which evolves
on two different time scales and its return distribution hence
regularly varies in time. This setting allows us to study the
interplay between the time scales and portfolio rebalancing.
Considering a risky asset with a lognormal return distribution
allows us to obtain an analytical form for the optimal
rebalancing period. This result is further generalized to other
stationary return distributions with finite variance and used to
explain some observations made for binary return distributions.
Our numerical simulations show that similar behavior can be
observed even for returns generated by the standard
$\textrm{GARCH}(1,1)$ process where consecutive returns are not
independent. Finally, we briefly study partial rebalancing where
the investor transfers only a certain part of the required
amount between cash and the risky asset. We show that this
strategy can enhance the long-term growth rate more than
intermittent rebalancing.

\section{Basic Model}
\label{sec:basic}
Consider a situation where an investor with an initial wealth
$W_0$ is allowed to repeatedly invest a fraction $f$ of the
current wealth to a risky asset while keeping the rest in cash.
We assume that the asset price $x(t)$ undergoes a multiplicative
stochastic process
\begin{equation}
\label{price1}
x(t+1)=\Big\{
\begin{aligned}
x(t)(1+r_1) & \qquad\text{with probability $\tfrac12+P_1$},\\
x(t)(1-r_1) & \qquad\text{with probability $\tfrac12-P_1$}
\end{aligned}
\end{equation}
at discrete time steps $t$ ($t=1,2,3,\dots$) and $x(1)=1$. Here
$r_1$ is a positive parameter ($0<r_1\leq 1$) representing the
rate of gain or loss of the investment, $\tfrac12+P_1$ is the
``winning'' probability and $P_1\in(0,\tfrac12]$ (when $P_1<0$, 
the asset is not profitable and it is advisable to refrain from
investment); it is assumed that they are both constant and known
to the investor.\footnote{Our parametrization based on
``excess'' winning probability $P_1$ is different from the
common one but it will prove very useful in later calculations
where it will allow us to obtain approximate results assuming
that $P_1$ is small.}
This ``symmetric'' setting can be easily generalized by assuming
distinct rates of gain/loss (\emph{e.g.}, $r_1$ and $r_1'$) as
well as their probabilities (\emph{e.g.}, $P_1$ and $P_1'$). To
keep the notation simple and to limit the number of parameters
to minimum, we treat only the symmetric case here. By setting
$r_1=1$, one recovers the original Kelly game studied
in~\cite{Kelly56}.

Since asset's properties do not change with time and investor's
wealth follows a multiplicative process, the investment fraction
set by a rational investor has to be the same in all time steps.
Investor's wealth after $N$ investment turns is therefore
\begin{equation}
\label{W_N}
W_N=W_0(1+fr_1)^w(1-fr_1)^{N-w}
\end{equation}
where $w$ and $N-w$ is the number of ``winning'' and ``loosing''
turns, respectively. Now we can introduce a so-called
exponential growth rate of investor's wealth, $G$, which is
defined by the relation $W_N=W_0\exp[GN]$. Its limit value has
the form
\begin{equation}
\label{G}
G:=\lim_{N\to\infty}\frac1N\ln\frac{W_N}{W_0}.
\end{equation}
One can easily show that for the given model parameters this
converges to the unique value
\begin{equation}
\label{Gsimple}
G(f)=(\tfrac12+P_1)\ln(1+fr_1)+(\tfrac12-P_1)\ln(1-fr_1).
\end{equation}
In the case of a general risky asset with return distribution
$\varrho(r)$, this formula generalizes to the form
\begin{equation}
\label{Ggeneral}
G(f)=\bavg{\ln(1+fr)}_{\varrho}
\end{equation}
where the average is over the return distribution $\varrho(r)$.
The long-term profitability of the risky asset can be measured
by the average return per time step, $R$. By definition,
$W_N=W_0(1+R_N)^N$ and $R=\lim_{N\to\infty}R_N$. Using
Eq.~(\ref{G}), $R$ can be expressed in terms of $G$ simply as
\begin{equation}
\label{R}
R=\exp(G)-1.
\end{equation}
Both $G$ and $R$ are functions of the asset parameters $r_1,P_1$
and of the investment fraction $f$.

According to the Kelly portfolio strategy~\cite{Kelly56}, for a
long-term investment it is best to maximize the growth rate $G$
(or, equivalently, the long-term return $R$)---this strategy is
therefore sometimes referred to as the growth-optimal investment
strategy. Starting from Eq.~(\ref{Gsimple}), simple computation
yields the optimal investment fraction
\begin{equation}
\label{fopt}
f_1^*=2P_1/r_1.
\end{equation}
Increasing the value of $P_1$ enhances the asset's profitability
and leads to an increased optimal investment fraction. On the
other hand, increasing $r_1$ enhances the asset's expected
return (when $P_1>0$) but it also increases the magnitude of
losses; overall it leads to a decreased value of $f^*$. When
$P_1>r_1$, we obtain $f^*>1$ which means that the investor
is advised to borrow additional money and invest them in the
risky asset too. When $P_1<0$ (the asset is not profitable),
$f^*<0$ which corresponds to the so-called short selling. For
simplicity we assume that both borrowing and short selling are
forbidden and hence $f\in[0;1]$.

\section{Transaction fees and intermittent portfolio rebalancing}
\label{sec:costs}
The requirement of keeping the investment fraction $f$ constant
implies that the investor needs to constantly rebalance the
portfolio: after a ``winning'' turn, some part of wealth has to
be moved from the asset to cash and after a ``loosing'' turn,
some additional wealth has to be invested in the asset. This
constant portfolio rebalancing may require payment of
substantial transaction fees. The question is, how the fees
affect the portfolio optimization process. In particular, we are
interested whether there are situations where the investor fares
better with intermittent rebalancing which is sub-optimal from
the point of view of the Kelly strategy but requires fewer money
transfers and hence lowers the transaction fees.

\subsection{Transaction fees}
We assume that for any wealth $X$ transferred from or to the
risky asset, a transaction fee $\alpha\abs{X}$ must be paid
($\alpha>0$; the absolute value reflects the fact that
transaction fees are paid regardless of the direction of the
transfer).\footnote{Since the investor's wealth grows without
bounds, the relative effect of any sub-linear fee
$\alpha\abs{X}^{\beta}$ is asymptotically zero in the long term.
The directly proportional fee $\alpha\abs{X}$ is hence the only
possible choice for the growth-optimal portfolio.}
How to include $\alpha$ in the derivation of the optimal
investment fraction presented above? Given that the portfolio is
properly balanced at a certain moment, the total amount invested
in the risky asset is $fW$. If the realized return from the
risky asset is $r$, the total wealth changes to $W(1+fr)$ and
the invested amount changes to $fW(1+r)$. If $r>0$, wealth $X>0$
needs to be transferred from the risky asset to cash to keep the
portfolio balanced. The resulting total wealth is then
$W(1+fr)-\alpha X$ and the invested amount is $fW(1+r)-X$. To
achieve the investment fraction $f$, it must hold that
\begin{equation}
\label{transfer-basic}
f\big[W(1+fr)-\alpha X\big]=fW(1+r)-X.
\end{equation}
From this formula it follows immediately that the total
transferred  volume is $X_{r>0}=Wrf(1-f)/(1-\alpha f)$. As
expected, no transfer is needed when $f=0$ or $f=1$; transaction
fees have no effect on portfolio optimization in these two
cases. When $r<0$, the transferred volume can be derived in a
similar way and has the form $X_{r<0}=Wrf(1-f)/(1-\alpha(1-f))$.
Now we know the wealth lost to transaction fees which allows us
to write investor's wealth after $N$ time steps
\begin{equation}
\label{W_N-fees}
W_N=W_0\left[1+fr_1-\frac{\alpha r_1f(1-f)}{1-\alpha f}\right]^w
\left[1-fr_1-\frac{\alpha r_1f(1-f)}{1-\alpha(1-f)}\right]^{N-w}.
\end{equation}
This is a generalization of Eq.~(\ref{W_N}) for the case with
transaction fees.

\begin{figure}
\centering
\includegraphics[scale=0.35]{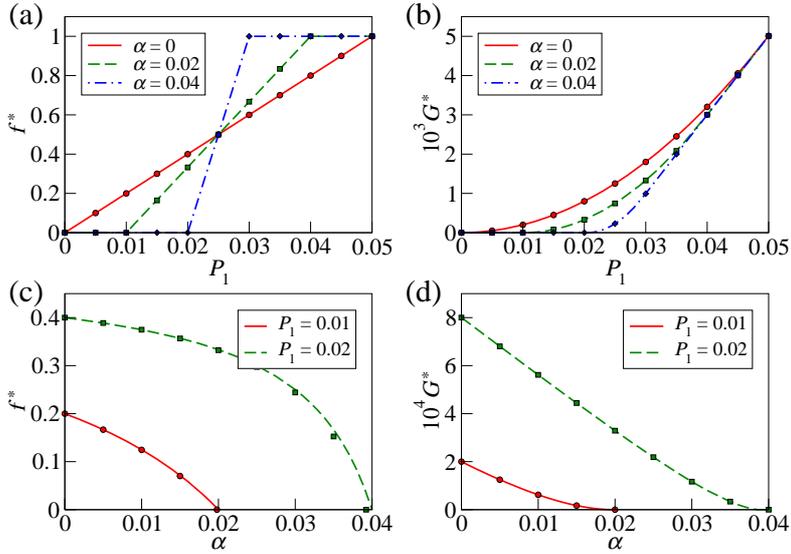}
\caption{The influence of transaction fees on the optimal
investment: the dependency on $P_1$ for $r_1,\alpha$ fixed (a,b)
and the dependency on $\alpha$ for $r_1,P_1$ fixed (c,d)
$r_1=10\%$ in all cases. Analytical and numerical results are
shown as lines and symbols, respectively.}
\label{fig:f_opt}
\end{figure}

It is straightforward to use Eq.~(\ref{W_N-fees}) to obtain the
exponential growth rate $G(f)$ and maximize it to get the
optimal investment fraction. Since the resulting quadratic
equation has complicated coefficients and provides little
insights to the behavior of the system, we introduce an
approximate approach which will be of great importance in later
more complicated cases. We expand $\dd G/\dd f$ in terms of
$\alpha,P_1,r_1$ and keep only terms up to order $\alpha$ (this
is motivated by the fact that the transaction fee coefficient
$\alpha$ is nowadays usually small in practice). Assuming that
$P_1$ and $r_1$ are sufficiently small, we neglect terms that
are of the order higher than $P_1^2$, $P_1r_1$, or $r_1^2$. The
resulting optimal fraction then has the simple form
\begin{equation}
\label{f_1}
f_1^*(\alpha)=\frac{2P_1-\alpha}{r_1-2\alpha}.
\end{equation}
Fig.~\ref{fig:f_opt} illustrates the dependency of this result
on both $P_1$ and $\alpha$. Naturally, in the limit $\alpha\to0$
we recover the fee-free result $f_1^*=2P_1/r_1$. Interestingly,
transaction fees may both decrease and increase the optimal
investment fraction (in comparison with the value corresponding
to $\alpha=0$). On the other hand, the average return is always
reduced by transaction fees.

Using Eq.~(\ref{f_1}), one can solve the equation
$f_1^*(\alpha)=0$ to obtain a lower bound for $P_1$ at which the
asset becomes profitable, $P_1^{\mathrm{low}}=\alpha/2$. As
expected, $P_1^{\mathrm{low}}$ is greater than the fee-free
lower bound which means that transaction fees decrease the
asset's profitability. Similarly, one can solve the equation
$f_1^*(\alpha)=1$ to obtain an upper bound for $P_1$ at which
the investor is advised to invest all wealth in the asset,
$P_2^{\mathrm{up}}=(r_1-\alpha)/2$ which is less than the
threshold $r_1/2$ valid for $\alpha=0$. We can conclude that
transaction fees narrow the region where non-trivial optimal
investment fractions ($0<f_1^*<1$) realize (this effect is well
visible in Fig.~\ref{fig:f_opt}a). Another point of view is that
transaction fees modify the optimal investment fraction $f^*$ so
that the transferred amounts (which are approximately
proportional to $(1-f)f$) are lowered. Transaction fees are in
this sense similar to friction in mechanics which also both
attenuates motion and leads to dissipation of energy (in the
case of transaction fees we face dissipation of wealth).

\subsection{Intermittent portfolio rebalancing}
\label{sec:binary}
While in the original Kelly game the investor should rebalance
the portfolio as often as possible (\emph{i.e.}, after each time
step), in the presence of transaction fees it may be profitable
to rebalance the portfolio less often. Our goal is to solve the
intermittent portfolio optimization problem first without and
then with transaction fees. Denoting the investor's rebalancing
period as $T$, the probability of ``winning'' in $w$ steps out
of $T$ is binomial and reads
$$
B_1(w\vert T)=
\binom{T}{w}(\tfrac12+P_1)^w(\tfrac12-P_1)^{T-w}.
$$
Since the asset's return in $T$ time steps can be written as
$$
r_w=(1+r_1)^w(1-r_1)^{T-w}-1,
$$
we know the return distribution and Eq.~(\ref{Ggeneral}) gives
the exponential growth rate
\begin{equation}
\label{GT}
G(f)=\sum_{w=0}^T B_1(w\vert T)\ln\big[1+fr_w\big]
\end{equation}
where substitution $T=1$ recovers $G(f)$ given by
Eq.~(\ref{Gsimple}). Using Eq.~(\ref{W_N-fees}), it is easy
to generalize this result to the case with both intermittent
rebalancing and transaction fees, yielding
\begin{equation}
\label{Grebalfees}
G(f)=\sum_{w=0}^T B_1(w\vert T)
\ln\bigg[1+fr_w-\frac{\alpha f(1-f)\abs{r_w}}
{1-\alpha\chi(f,r_w)}\bigg]
\end{equation}
where $\chi(f,r_w)=f$ if $r_w>0$ and $\chi(f,r_w)=1-f$
otherwise.

Eq.~(\ref{Grebalfees}) cannot be maximized analytically in
general and one has to resort to numerical techniques. When
$T=2$, the approach that we developed to derive Eq.~(\ref{f_1})
yields
\begin{equation}
\label{f_2}
f_2^*(\alpha)=\frac{8P_1-\alpha(2+r_1)}{4r_1-2\alpha(2+r_1)}.
\end{equation}
Notice that in the limit $\alpha\to0$, this result is identical
with the optimal portfolio fraction for rebalancing after each
turn which is a direct consequence of assuming that $P_1$ and
$r_1$ are small.\footnote{In a general case, $f_2^*(0)$ may be
considerably different from $f_1^*(0)$. For our setting, for
example, one can find the approximate result
$f_2^*(0)\approx\tfrac{2P_1}{r_1}\big(
1-\tfrac12r_1^2+3r_1P_1-4P_1^2\big)$ which shows that $f_2^*(0)$
is indeed different from $f_1^*(0)=2P_1/r_1$.}
Numerical tests show that Eq.~(\ref{f_2}) is reasonably precise
for $\alpha,P_1\ll 1$.

\begin{figure}
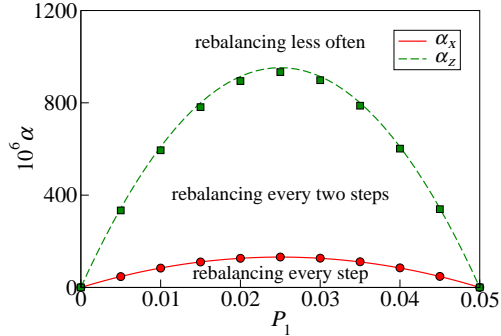

\centering
\onepic{a_vs_P1-r=10}
\caption{Intermittent rebalancing: numerical and approximate
analytical values (shown as symbols and lines, respectively) of
$\alpha_x$ and $\alpha_z$ for $r_1=10\%$.}
\label{fig:two}
\end{figure}

Solution of the optimization problem for $T=2$ allows us to ask
what transaction fee $\alpha_x$ makes rebalancing every other
turn as profitable (in terms of the exponential growth rate $G$)
as rebalancing in every turn. Using Eqs.~(\ref{fopt}),
(\ref{GT}), (\ref{f_2}) one can show that when $\alpha=0$, the
difference of the optimal growth rates per turn is
$$
G_1^*(0)-\tfrac12 G_2^*(0)=P_1^2(r_1-2P_1)^2
$$
where we neglected fifth and higher powers of $P_1$ and $r_1$ in
the result. (The factor $1/2$ at $G_2^*(0)$ converts the
exponential growth rate in two-turn basis to the growth rate per
turn.) Assuming that $P_1,r_1,\alpha$ are small, it is also
possible to find that the growth rates depend on $\alpha$ as
\begin{align*}
G_1^*(\alpha)&=G_1^*(0)-\frac{2P_1}{r_1}
(r_1-2P_1)\alpha+O(\alpha^2),\\
G_2^*(\alpha)&=G_2^*(0)-\frac{P_1}{r_1}(r_1-2P_1)(2+r_1)\alpha+
O(\alpha^2).
\end{align*}
Both growth rates are for $P_1=0$ and $P_1=r_1/2$ independent of
$\alpha$. This is not surprising: in those cases is $f^*=0$ or
$f^*=1$ and hence no rebalancing is necessary and the optimal
exponential growth rate is unaffected by transaction fees.
Combining the obtained results together, the equality
$G_1^*(\alpha)=G_2^*(\alpha)/2$ can be solved with respect to
$\alpha$, leading to
\begin{equation}
\label{ax}
\alpha_x=2r_1P_1\frac{r_1-2P_1}{2-r_1}
\end{equation}
which represents the magnitude of $\alpha$ for which rebalancing
in every turn and in every other turn are equally profitable.
As shown in Fig.~\ref{fig:two}, this formula is very accurate
even for moderate values of parameters $P_1,r_1$. It is
instructive to note that the threshold fee value $\alpha_x$ is
small for weakly profitable assets ($P_1$ small) and in
particular for assets with small return in one step ($r_1$
small).

In a very similar way it is possible to study the transaction
fee at which rebalancing every two turns is as profitable as
rebalancing every three turns. Interestingly, the resulting
value $\alpha_y=2P_1(r_1-2P_1)$ is for $r_1<1$ greater than
$\alpha_x$ (by the factor of $(2-r_1)/r_1$). This means that
rebalancing every three turns is quite ineffective and hence it
is meaningful to ask what $\alpha_z$ makes rebalancing every two
and four turns equally profitable. The corresponding value
\begin{equation}
\label{az}
\alpha_z=16P_1r_1\frac{r_1-2P_1}{2+r_1} 
\end{equation}
is greater than $\alpha_x$ for $\alpha<9/14$ and it is smaller
than $\alpha_y$ for $r_1<2/7$. This means that rebalancing every
three turns is sub-optimal in the case of small investment
returns: it is better to rebalance either more
(for $\alpha>\alpha_z$) or less ($\alpha<\alpha_z$) often. As
shown in Fig.~\ref{fig:two}, while precision of $\alpha_z$ is
lower than that of $\alpha_x$, obtained values agree well with a
purely numerical treatment of the problem.\footnote{For the sake
of completeness, the optimal investment fractions for
rebalancing every three and four turns are
$f_3^*=(2P_1-\alpha/2)/(r_1-2\alpha)$ and
$f_4^*=[32P_1-3\alpha(2+r_1)]/[16r_1-6\alpha(2+r_1)]$,
respectively, while the optimal exponential growth rates are
$G_3^*(\alpha)=G_3^*(0)-3(r_1-2P_1)P_1/r_1+
O(\alpha^2)$ and $G_4^*(\alpha)=G_4^*(0)-\tfrac32
P_1(r_1-2P_1)(2+r_1)/r_1+O(\alpha^2)$, respectively.}

\begin{figure}
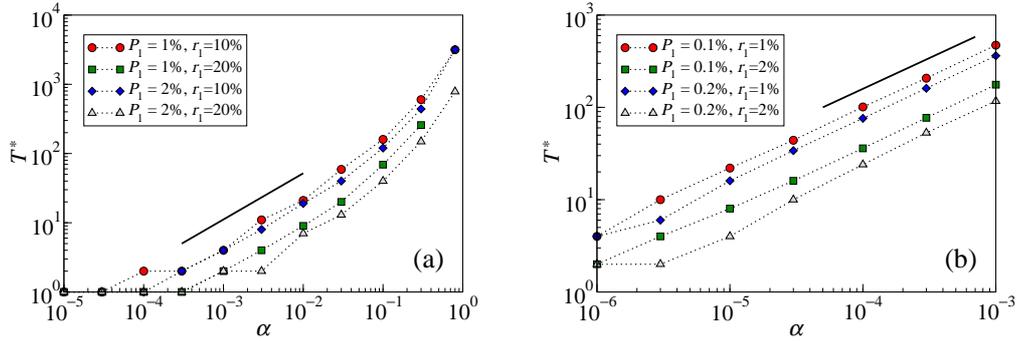

\centering
\onepic{T_vs_fee-large}\qquad
\onepic{T_vs_fee-small}
\caption{Optimal rebalancing periods $T^*$ vs $\alpha$ for
different choices of $r_1$ and $P_1$; the indicative thick lines
have slope $2/3$.}
\label{fig:Topt}
\end{figure}

When $r_1,P_1,\alpha$ are given, it is natural to ask what
rebalancing period $T^*$ maximizes the exponential growth rate
per turn. While this question cannot be answered analytically,
it is straightforward to solve it numerically. Results are shown
in Fig.~\ref{fig:Topt} for various choices of $r_1,P_1$. As can
be seen, $T^*$ decreases with both $P_1$ and $r_1$. This agrees
with the growth of the threshold values $\alpha_x,\alpha_z$
with $P_1$ (until $P_1<r_1/4$) and $r_1$ (see Eqs.~(\ref{ax}),
(\ref{az})). When transaction fees are small, $T^*$ is
proportional to $\alpha^{2/3}$---a behavior which will be
explained in Sec.~\ref{sec:lognorm}. When
$\alpha\gtrsim10^{-2}$, this scaling breaks down and $T^*$ grows
even faster than linearly. Since this mode of behavior occurs
only for exceedingly large transaction fees (note that
$\alpha=1$ corresponds to confiscating all invested amount), we
do not study it further.

\subsection{Risky assets with multiple time scales}
\label{sec:scales}
Assets' properties are in real life generally non-stationary. To
analyze investment in an asset with time-varying properties, we
propose a simple model where the price of the asset undergoes a
stochastic binary process on two distinct time scales. In
addition to the basic time scale $1$, we add a longer scale of
length $T_2$. We assume that price of the asset undergoes a
multiplicative dynamics given by Eq.~(\ref{price1}) at all time
steps and when $(t \mod T_2)=0$, there is an additional return
$\pm r_2$ with probabilities $\tfrac12+P_2$ and $\tfrac12-P_2$,
respectively (as before, asset parameters are constrained to
$0<r_2\leq1$ and $0<P_2\leq 1/2$). This framework is a simple
generalization of the original Kelly game to the case with
non-stationary game properties and multiple time scales.

\begin{figure}
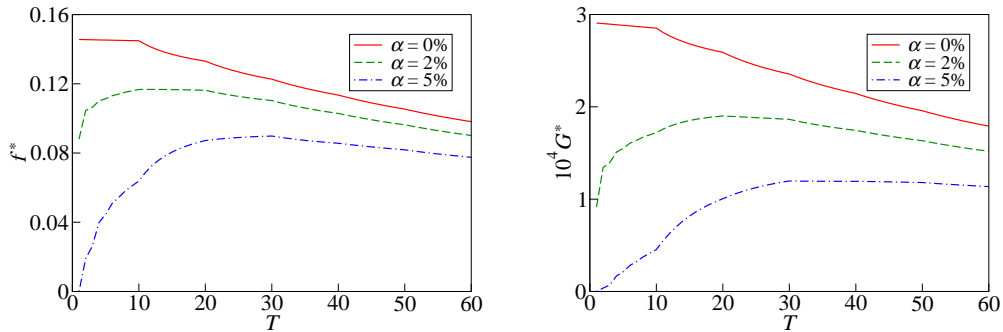

\centering
\onepic{fopt-case_2}\qquad\onepic{Gopt-case_2}
\caption{Optimal investment fraction and growth rate of the
Kelly portfolio for an asset with price change on two time
scales with $P_1=-0.01$, $r_1=0.05$, $P_2=0.05$, $R=0.5$,
$T_2=10$.}
\label{fig:two_time_scales}
\end{figure}

The simplest case is when the investor  keeps the investment
fraction $f$ constant and rebalances the investment every $T$
steps. Since price dynamics is still binary, we can parametrize
the outcome by the number of ``winning'' turns on the basic time
scale, $w_1$, and by the number of ``winning'' turns on the
longer time scale, $w_2$. While $w_1$ is simply constrained to
$0,\dots,T$, the upper bound for $w_2$ can be either
$T\backslash T_2$ or $1+T\backslash T_2$ ($\mymod$ and
$\backslash$ denote the modulus operator and integer division,
respectively). Simple algebra shows that the odds of the two
cases are $1-(T\mymod T_2)/T_2$ and $(T\mymod T_2)/T_2$,
respectively, hence we can write the long-term exponential
growth rate of the portfolio in the form
\begin{equation}
\label{monster}
\begin{aligned}
G=&\big(1-\tfrac{T\mymod T_2}{T_2}\big)
\sum_{w_1=0}^T\sum_{w_2=0}^{t} B_1(w_1\vert T)
B_2(w_2\vert t)\ln\bigg[1+
fr_t-\frac{\alpha f(1-f)\abs{r_t}}{1-\alpha\chi(f,r_t)}\bigg]+\\
&+\tfrac{T\mymod T_2}{T_2}
\sum_{w_1=0}^T\sum_{w_2=0}^{t+1} B_1(w_1\vert T)
B_2(w_2\vert t+1)\ln\bigg[1+fr_{t+1}-
\frac{\alpha f(1-f)\abs{r_{t+1}}}{1-\alpha\chi(f,r_{t+1})}\bigg]
\end{aligned}
\end{equation}
where $t:=T\backslash T_2$,
$$
r_t=(1+r_1)^{w_1}(1-r_1)^{T-w_1}(1+r_2)^{w_2}(1-r_2)^{t-w_2}-1
$$
is the compound return before transaction fees are applied
and $B_2(w_2\vert t)$ is the binomial probability of $w_2$ wins
in $t$ trials when the winning probability is $\tfrac12+P_2$.
Albeit principally simple, the described situation is out of
scope of analytical optimization tools and hence we report only
numerical results here. The most interesting behavior is
obtained when the risky asset is profitable only on the longer
time scale (that is, $P_1<0$ and $P_2>0$). The need to rebalance
often (which is a principal property of the Kelly portfolio)
then directly competes with the asset profitability on a longer
time scale. An example of the resulting behavior is shown in
Fig.~\ref{fig:two_time_scales} where irregularities
corresponding to the longer time scale are visible on both
$f^*(T)$ and $G^*(T)$. On the other hand, when $T\gg T_2$, the
two time scales merge into average behavior of the risky asset
and the irregularities are not visible anymore. We can conclude
that the presence of multiple time scales is important only if
portfolio rebalancing occurs in time intervals comparable to the
longest time scale of asset's returns.

\section{Intermittent rebalancing for lognormal returns}
\label{sec:lognorm}
Now we shall study portfolio optimization for a simple asset
with lognormally distributed returns. We assume that the asset's
price $p_i(t)$ ($i=1,\dots,N$) undergoes an uncorrelated
multiplicative random walk
$$
p(t)=p(t-1)\ee^{\eta(t)}
$$
where random variable $\eta(t)$ is drawn from the Gaussian
distribution with mean $m$ and variance $D$. Consequently,
returns of the asset have the form
$$
r(t):=\frac{p(t)}{p(t-1)}-1=\ee^{\eta(t)}-1.
$$
Using the same notation as above, the investor's expected
exponential growth rate has the form
$G(f)=\avg{\ln(1+fr)}$ where the average is over different
values of $\eta$. Written in detail, the previous expression
reads
$$
G(f)=\int_{-\infty}^{\infty}\dd\eta\varrho(\eta)
\ln\big[1+f(\ee^{\eta}-1)\big]
$$
where $\varrho(\eta)$ is the Gaussian probabilistic density of
returns. With transaction fees, $G(f)$ generalizes to the form
\begin{equation}
\label{lognorm-Gf}
G(f)=\int_{-\infty}^{\infty}\dd\eta\varrho(\eta)
\ln\big[1+f(\ee^{\eta}-1)-\frac{\alpha f(1-f)\abs{\ee^{\eta}-1}}
{1-\alpha\chi(f,\ee^{\eta}-1)}\big].
\end{equation}
When $\alpha=0$, it is known (see~\cite{LaMeZh10}) that the
optimal investment fraction has the approximate form
\begin{equation}
\label{lognorm-f0}
f_0^*(m,D)=\frac12+\frac{m}{D}
\end{equation}
which is valid when $m,D\ll1$. Here $f_0^*=0$ for $m<-D/2$ and 
$f_0^*=1$ for $m>D/2$ (when $f$ is out of the range $[0,1]$, the
investor has a non-zero probability of going bankrupted and
hence the long-term growth rate is automatically
zero~\cite{LaMeZh10}). In our following analysis we will hence
assume that $m$ and $D$ are of the same order of smallness.

\begin{figure}
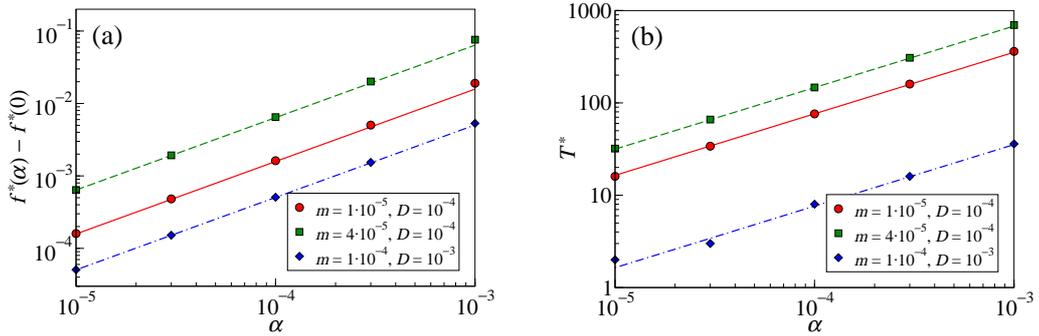

\centering
\onepic{df_vs_a-lognorm}\qquad\onepic{Topt_vs_a-lognorm}
\caption{The dependency of the optimal investment fraction (a)
and the optimal rebalancing period (b) on $\alpha$: numerical
and analytical results are shown with symbols and lines,
respectively.}
\label{fig:lognorm}
\end{figure}

When $\alpha>0$, we search the optimal fraction in the form
$f^*(\alpha)=f^*(0)+u$ where the correction $u$ is small when
$\alpha$ is small. Since our goal is to find the highest order
correction to $f^*$, we neglect the term
$\alpha\chi(f,\ee^{\eta}-1)$ in Eq.~(\ref{lognorm-Gf}). The
optimal investment fraction is the solution of
$\partial G/\partial f=0$. By exchanging the order of derivation
and  integration we obtain
$$
\frac{\partial G}{\partial f}=\int_{-\infty}^0
\frac{(\ee^{\eta}-1)(1+\alpha-2\alpha f)\varrho(\eta)\dd\eta}
{1+f(\ee^{\eta}-1)[1+\alpha(1-f)]}+\int_0^{\infty}
\frac{(\ee^{\eta}-1)(1-\alpha+2\alpha f)\varrho(\eta)\dd\eta}
{1+f(\ee^{\eta}-1)[1-\alpha(1-f)]}
$$
where it was necessary to write two separate terms because of
the absolute value $\abs{\ee^{\eta}-1}$ present in $G(f)$. We
can now substitute $f=f_0^*+u$ where $f_0^*$ is the solution of
$\partial G/\partial f=0$ for $\alpha=0$ (see
Eq.~(\ref{lognorm-f0}) above). Assuming that both $\alpha$ and
$u$ are small, the integrand of the first integral can be
approximated as
$$
\frac{(1+\alpha-2\alpha f_0^*)x(\eta)}
{1+(u+\alpha f_0^*(1-f_0^*)x(\eta)}
\approx x(\eta)\big[1+\alpha(1-2f_0^*)-
\big(u+\alpha f_0^*(1-f_0^*)\big)x(\eta)\big]
$$
where $x(\eta)=(\ee^{\eta}-1)/(1+(\ee^{\eta}-1)f_0^*)$. The
second integral can be manipulated in a similar way; by putting
the results together we  get
$$
\frac{\partial G}{\partial f}=\int_{-\infty}^{\infty}
\dd\eta\varrho(\eta)\big[x(\eta)+\alpha(2f_0^*-1)\abs{x(\eta)}-
u x(\eta)^2\big]
$$
which is equivalent to three separate integrals. The first one
is zero by definition (we assume that $f_0^*$ is the solution
for $\alpha=0$). For the second and third integral, we use
$x(\eta)\approx\ee^{\eta}-1$ (because $m,D\ll1$ and hence $\eta$
is small) and
$\varrho(\eta)\approx\exp[-\eta^2/(2D)]/\sqrt{2\pi D}$ (because
$D\ll1$ and $\abs{m}\leq D/2$ and hence $\abs{m}\ll\sqrt{D}$).
While the integration results are complicated and involve the
error function, for $D\ll1$ we can simplify them further to
finally obtain
$$
\frac{\partial G}{\partial f}=
\alpha(2f_0^*-1)D-u\sqrt{\pi D^3/2}.
$$
Thus $u$ that maximizes $G$ (solution of
$\partial G/\partial f=0$) has the form
$$
u=\alpha m\sqrt{\frac8{\pi D^3}}
$$
with the next contributing term of the order of $O(\sqrt{D})$.
In combination with Eq.~(\ref{lognorm-f0}) we have
\begin{equation}
\label{lognorm-correction}
\begin{aligned}
f^*(m,D,\alpha)&=\frac12+\frac{m}{D}+
\alpha m\sqrt{\frac8{\pi D^3}},\\
G^*(m,D,\alpha)&=G^*(m,D,0)-\alpha\bigg(
\frac14-\frac{m^2}{D^2}\bigg)\sqrt{\frac{2D}{\pi}}.
\end{aligned}
\end{equation}
As shown in Fig.~\ref{fig:lognorm}a, this agrees well with
numerical results for $f^*$ (numerical results for $G^*$ are not
shown).

When $\alpha=0$, by expanding $\ln[1+f(\ee^{\eta}-1)]$ in
Eq.~(\ref{lognorm-Gf}) into a series of $f(\ee^{\eta}-1)$ we
get the following approximate expression for the optimal
exponential growth rate
$$
G^*(m,D)=\frac D2\bigg(\frac12+\frac mD\bigg)^2-
\frac{D^2}4\bigg(\frac14-\frac{m^2}{D^2}\bigg)^2+O(D^3).
$$
When the rebalancing period is $T$, the compound return of the
asset is again lognormally distributed, this time with $\eta$
drawn from the Gaussian distribution with mean $Tm$ and variance
$TD$ (here we take the advantage from the fact that the Gaussian
distribution is stable). Using the above expression for
$G^*(m,D)$ we can write the resulting optimal growth rate per
time step as
\begin{equation}
G^*(m,D,T):=G^*(Tm,TD)/T\approx
\frac D2\bigg(\frac12+\frac mD\bigg)^2-
\frac{TD^2}4\bigg(\frac14-\frac{m^2}{D^2}\bigg)^2
\end{equation}
which is a decreasing function of $T$ as expected. Combining
this result with Eq.~(\ref{lognorm-correction}) produces a
general dependency of the optimal growth rate on $T$ and
$\alpha$. This dependency can be simply maximized with respect
to $T$, yielding
\begin{equation}
\label{lognorm-Topt}
T^*(m,D,\alpha)=\frac{\alpha^{2/3}}{D}\sqrt{\frac8{\pi}}
\bigg(\frac14-\frac{m^2}{D^2}\bigg)^{-2/3}
\end{equation}
which is confirmed by comparison with numerical simulations in
Fig.~\ref{fig:lognorm}b (small irregularities visible for
$D=10^{-3}$ are caused by true $T^*$ being an integer number).
After multiplying Eq.~(\ref{lognorm-Topt}) with $D$ we obtain an
expression for $DT^*:=D^*$ which can be understood as an optimal
variance of lognormally distributed returns. When $\alpha=0$,
this optimal variance is zero, indicating that the investor
should rebalance the portfolio continuously.

\begin{figure}
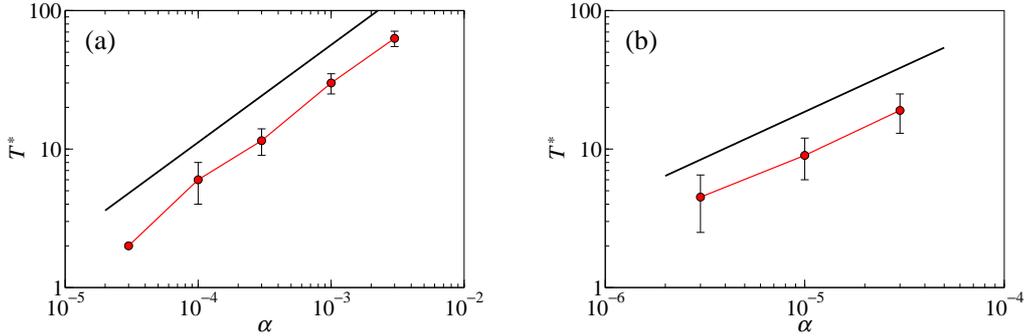

\centering
\onepic{Topt_vs_fee-Student}\qquad\onepic{Topt_vs_fee-GARCH}
\caption{Optimal rebalancing period $T^*$ vs $\alpha$ for: (a)
Student-based returns and (b) GARCH-based returns. The
indicative dashed lines have slopes $0.70$ and $0.66$,
respectively.}
\label{fig:Topt2}
\end{figure}

Results derived for the lognormal distribution of returns are of
particular importance when intermittent rebalancing is
considered. If we write the return at time $t$ as
$r(t):=p(t)/p(t-1)-1=\ee^{\varrho(t)}-1$ where values
$\varrho(t)$ are drawn from a probabilistic distribution with
finite mean and variance, the compound return over a period of
$T$ turns is
$$
r_T(t):=p(t)/p(t-T)-1=
\exp\bigg[\sum_{u=0}^{T-1}\varrho(t-u)\bigg]-1.
$$
According to the central limit theorem, if variables
$\varrho(t)$ are independent and $T$ is large, the sum
$\sum_{u=0}^{T-1}\varrho(t-u)$ is approximately normally
distributed and hence compound return $r_T(t)$ follows a
lognormal distribution when the rebalancing period $T$ is long.
This immediately explains the scaling $T^*\sim\alpha^{2/3}$
which was found numerically for binary returns in
Sec.~\ref{sec:binary}.\footnote{When the random variable
$\varrho(t)$ has two possible values $\ln(1\pm r_1)$ with
probabilities $1/2\pm P_1$, respectively, one recovers the
binary returns studied in Sec.~\ref{sec:binary}.}
The same reasoning applies to any $\varrho(t)$ following a broad
distribution with finite variance. As an example, we use returns
$r(t)=\ee^{\sigma\chi(t)}-1$ where $\chi(t)$ is Student's
distribution with two degrees of freedom (the tails of $\chi(t)$
then decay as $\chi^{-3}$, hence $\chi(t)$ has finite variance).
Since Student's distribution is not stable, the distribution of
returns for an arbitrary rebalancing period $T$ does not have a
closed form and one cannot attempt to find the optimal
rebalancing period analytically. We employ numerical simulations
in which the exponential growth rate is maximized with respect
to the investment fraction $f$ over $10^6$ time steps for
rebalancing periods in the range $1,\dots,100$. The resulting
optimal growth rates $G^*(T)$ are averaged over $10^3$
independent realizations of returns and finally yield the
optimal rebalancing period which is again roughly proportional
to $\alpha^{2/3}$ (see Fig.~\ref{fig:Topt2}a).

When aiming at even more realistic return distributions, it is
a question whether $\alpha^{2/3}$-scaling holds for returns
with some degree of dependence (memory). Since there are various
central limit theorems for dependent variables
\cite{HoeRob48,ProSta00}, one expects that when the dependence
of returns is sufficiently weak, above-obtained results continue
to hold. This is confirmed by our simulations with returns
generated by a $\textrm{GARCH}(1,1)$
process~\cite{Engle01,Taylor08} with parameters
$\alpha_0=10^{-5}$, $\alpha_1=0.2$, $\beta=0.7$ (these parameter
values are similar to those inferred from S\&P index data
in~\cite{Duan95}). The optimal rebalancing period---obtained by
the same simulation approach as above for the Student-based
returns---is again proportional to $\alpha^{2/3}$ (see
Fig.~\ref{fig:Topt2}b). This confirms that our main result is
highly robust with respect to the nature of the return
distribution. Detailed insights on the degree of dependence at
which this scaling breaks down are however yet to come. 

\section{Partial rebalancing}
\label{sec:partial}
So far we only considered mitigating the impact of transaction
fees by intermittent rebalancing. There is another approach,
which we call \emph{partial rebalancing}, where only part of the
required amount is transferred between cash and the asset. The 
transferred amount required to keep the investment fraction
fixed is represented by $X$ in Eq.~(\ref{transfer-basic}). If
only part $\varepsilon$ of the required capital is transferred
($\varepsilon\in(0,1]$ is a rebalancing parameter),
$1/\varepsilon$ steps would be needed to transfer the whole.
Hence one can expect that partial rebalancing with $\varepsilon$
should be similar to intermittent rebalancing with period
$T\approx 1/\varepsilon$. Since partial rebalancing is
parametrized by a continuous parameter $\varepsilon$, it allows
for smoother setting of the portfolio than intermittent
rebalancing where the rebalancing period is integer.

\begin{figure}
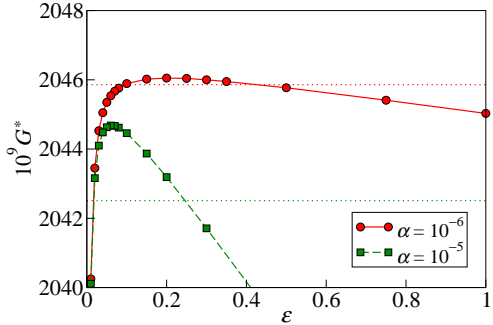

\centering
\onepic{partial_rebalancing}
\caption{Optimal growth rate $G^*$ vs $\varepsilon$ for binary
returns with $P_1=0.1\%$ and $r_1=1\%$. The two thick dotted
lines show performance of intermittent rebalancing for
$\alpha=10^{-6}$, $T=4$ (upper) and $\alpha=10^{-5}$, $T=22$
(lower), respectively.}
\label{fig:partial}
\end{figure}

When $\varepsilon<1$, the desired investment fraction is almost
never achieved by partial rebalancing and the actual invested
fraction fluctuates around it. (The smaller the value of
$\varepsilon$, the larger the deviations; when $\varepsilon=1$,
the standard rebalancing is recovered and the stake is always
adjusted accurately.) Due to these irregularities, partial
rebalancing is less accessible to analytical treatment and we
present only numerical results here. As shown in
Fig.~\ref{fig:partial}, the optimal growth rates are achieved for
$\varepsilon$ inside the range $(0,1]$. These rates outperform
the optimal values obtained with intermittent rebalancing for
both studied values of $\alpha$. If growth rate improvement is
measured in respect to standard rebalancing (the same as obtained
with $\varepsilon=1$), improvements obtained with partial
rebalancing are 23\% (for $\alpha=10^{-6}$) and 18\% (for
$\alpha=10^{-5}$) better than those obtained with intermittent
rebalancing. As foreseen above, optimal values of $\varepsilon$
($0.21$ and $0.06$, respectively) approximately correspond to
the optimal periods of intermittent rebalancing ($4$ and $22$,
respectively).

\section{Discussion}
While transaction fees represent an important factor limiting
investor's profit, in finance literature they are often
considered as uninteresting and neglected in order to keep the
analysis simple and focused. However, money transfers required
by active portfolio optimization strategies may be considerable
and the effect of transaction fees significant. In this work we
investigated this effect on the growth-optimal/Kelly portfolio
in detail. To this end we studied a toy risky asset with a
binary return distribution, an asset with time-depending return
distribution, and more realistic assets with lognormal and
fat-tailed return distributions. Our results show that
transaction fees indeed have substantial impact on investment
profitability, in particular when the average return of the
risky asset is low. Their influence is greatest when the
investment fraction is $1/2$. This is natural because the wealth
volumes transferred in rebalancing are proportional to $f(1-f)$
and hence they are maximized at $f=1/2$.

We showed for various settings that when the transaction fee
coefficient $\alpha$ is sufficiently high, for the investor it
may be more profitable to adjust the portfolio less frequently
and an optimal rebalancing period $T^*$ arises. In the case of a
lognormal distribution of returns, the optimal optimal
rebalancing period was analytically shown to be proportional to
$\alpha^{2/3}$ for small $\alpha$. When $\alpha$ is small yet
$T^*$ is sufficiently long for the central limit theorem to be
an appropriate approximation, the optimal rebalancing period
scales with $\alpha^{2/3}$ for any independent returns drawn
from a distribution with finite mean and variance. Our numerical
simulations confirm this for binary returns, returns based on
Student's distribution, and even for returns with memory
modeled by a $\textrm{GARCH}(1,1)$ process where the requirement
of independence is violated. We showed that a so-called partial
rebalancing can also reduce the impact of transaction fees and
hence improve the performance of the Kelly strategy.

Besides presented results, several research questions remain
open. Firstly, while transaction fees are maximized by $f=1/2$
when investing in one asset, the situation gets more complicated
when investment is distributed among several assets. That
situation can be further generalized by assuming correlated
asset returns. Secondly, through the paper we have assumed that
parameters of the return distribution are known to the investor.
Investment optimization hence only consists of choosing the
right investment fraction. In real life, the return distribution
itself is unknown and its estimation is part of the optimization
process. Whether the presented results hold also this case is an
open question. Finally, results for partial rebalancing
presented in Sec.~\ref{sec:partial} show that this can be a
superior approach to the Kelly optimization in presence of
transaction fees. Observed similarity between optimal values of
rebalancing parameters $T$ and $\varepsilon$ suggests that many
of analytical results found here for intermittent rebalancing
may hold also for partial rebalancing. Verification of this
hypothesis remains as an important challenge for future
research.

\section*{Acknowledgements}
This work was partially supported by the Shanghai Leading
Discipline Project (grant no. S30501). We acknowledge helpful
comments of our anonymous reviewers.


\begin{thebibliography}{99}
\bibitem{EGBG06} E. J. Elton, M. J. Gruber, S. J. Brown, W. N.
Goetzmann, Modern Portfolio Theory and Investment Analysis, 7th
edn., Wiley, 2006.

\bibitem{ZZ06} S. Zenios, W. Ziemba, Eds.,
Handbook of Asset and Liability Management, Volume 1,
North-Holland, 2006.

\bibitem{Mark52} H. M. Markowitz, The Journal of Finance 7,
77--91, 1952.

\bibitem{Kelly56} J. L. Kelly Jr.,
Bell System Technical Journal 35, 917--926, 1956.
% A New Interpretation of Information Rate

\bibitem{Mark76} H. M. Markowitz, The Journal of Finance 31,
1273--1286, 1976.
% Investment for the Long Run: New Evidence for an Old Rule

\bibitem{MaMaZh98} M. Marsili, S. Maslov, Y.-C. Zhang, Physica A
253, 403--418, 1998.
% Dynamical Optimization Theory of a Diversified Portfolio

\bibitem{LaMeZh10} P. Laureti, M. Medo, Y. C. Zhang,
Quantitative Finance 10, 689--697, 2010.
% Analysis of Kelly-optimal portfolios

\bibitem{BrWh96} S. Browne, W. Whitt,
Advances in Applied Probability 28, 1145--1176, 1996.

\bibitem{MePiZh08} M. Medo, Y. M. Pis'mak, Y.-C. Zhang,
Physica A 387, 6151--6158, 2008.

\bibitem{FerKar09} E. R. Fernholz, I. Karatzas,
in Handbook of Numerical Analysis. Mathematical Modeling and
Numerical Methods in Finance, A. Bensoussan, Ed., Elsevier,
89--168, 2009.
% Stochastic portfolio theory: A survey

\bibitem{Slanina99} F. Slanina, Physica A 269, 554--563, 1999.
% On the possibility of optimal investment

\bibitem{MLZiBl92} L. C. MacLean, W. T. Ziemba, G. Blazenko,
Management Science 38, 1562--1585, 1992.

\bibitem{GroZho93} S. J. Grossman, Z. Zhou,
Mathematical Finance 3, 241--276, 1993.

\bibitem{Thorp06} E. O. Thorp, in Handbook of Asset and
Liability Management, Volume 1, S. Zenios and W. Ziemba, Eds.,
North-Holland, 385--428, 2006.

\bibitem{YaSo09} G. Yaari, S. Solomon,
EPJ B 73, 625--632, 2010.
% Cooperation Evolution in Random Multiplicative Environments

\bibitem{MLeZiem06} L. C. MacLean, W. T. Ziemba, 
in Handbook of Asset and Liability Management, Volume 1,
S. Zenios and W. Ziemba, Eds., North-Holland, 429--474, 2006.

\bibitem{MoCha10} D. Morton de Lachapelle, D, Challet,
New Journal of Physics 12, 075039, 2010.
% Turnover, account value and diversification of real traders:
% evidence of collective portfolio optimizing behavior

\bibitem{Merton90} R. C. Merton, Continuous-Time Finance,
Blackwell, 1990.

\bibitem{MasZha98} S. Maslov, Y. C. Zhang,
International Journal of Theoretical and Applied Finance 1,
377--387, 1998.
% Optimal Investment Strategy for Risky Assets

\bibitem{HoeRob48} W. Hoeffding, H. Robbins,
Duke Mathematical Journal 15, 773--80, 1948.
% The central limit theorem for dependent variables

\bibitem{ProSta00} Yu.V. Prokhorov, V. Statulevicius, Eds.,
Limit Theorems of Probability Theory, Springer, 2000.

\bibitem{Engle01} R. Engle,
Journal of Economic Perspectives 15, 157--168, 2001.
% GARCH 101: The Use of ARCH/GARCH Models in Applied Econometrics

\bibitem{Taylor08} S. J. Taylor,
Modelling financial time series, 2nd Ed., World Scientific, 2008.

\bibitem{Duan95} J.-C. Duan,
Mathematical Finance 5, 13--32, 1995.
% The Garch Option Pricing Model
\end{thebibliography}
\end{document}